\documentclass{article}

\usepackage{PRIMEarxiv}

\usepackage[utf8]{inputenc} 
\usepackage[T1]{fontenc}    
\usepackage{hyperref}       
\usepackage{url}            
\usepackage{booktabs}       
\usepackage{amsfonts}       
\usepackage{nicefrac}       
\usepackage{microtype}      
\usepackage{lipsum}
\usepackage{fancyhdr}       
\usepackage{graphicx}       

\usepackage{cite}
\usepackage{amsmath,amssymb,amsfonts}
\usepackage{algorithmic}
\usepackage{graphicx}
\usepackage{textcomp}
\usepackage{xcolor}

\usepackage{graphicx, pdfpages} 
\usepackage{caption}
\usepackage{subcaption} 
\usepackage[raggedrightboxes]{ragged2e}
\usepackage{array}
\usepackage[frozencache]{minted}
\usepackage{listings}
\usepackage{mdframed, hyperref, comment}

\def\BibTeX{{\rm B\kern-.05em{\sc i\kern-.025em b}\kern-.08em
    T\kern-.1667em\lower.7ex\hbox{E}\kern-.125emX}}

\makeatletter
\newcommand{\linebreakand}{%
  \end{@IEEEauthorhalign}
  \hfill\mbox{}\par
  \mbox{}\hfill\begin{@IEEEauthorhalign}
}
\makeatother

\title{Icing on the Cake: Automatic Code Summarization at Ericsson
\thanks{Accepted at the \href{https://conf.researchr.org/details/icsme-2024/icsme-2024-industry-track/3/Icing-on-the-Cake-Automatic-Code-Summarization-at-Ericsson}{2024 International Conference on Software Maintenance and Evolution (ICSME) 2024 - Industry Track}
}
}
\author{
  Giriprasad Sridhara, Sujoy Roychowdhury, Sumit Soman, Ranjani H G \\
  Ericsson, GAIA \\
  Bangalore, India\\
  \texttt{\{giriprasad.sridhara, sujoy.roychowdhury, sumit.soman, ranjani.h.g\}ericsson@com} \\
   \And
  Ricardo Britto \\
  Ericsson, SA BOS  \& Blekinge Institute of Technology\\
Karlskrona, Sweden \\
  \texttt{ricardo.britto@ericsson.com} \\
}

\begin{document}
\maketitle

\begin{abstract}
This paper presents our findings on the automatic summarization of Java methods within Ericsson, a global telecommunications company. We evaluate the performance of an  approach called Automatic Semantic Augmentation of Prompts (ASAP), which uses a Large Language Model (LLM) to generate leading summary comments (Javadocs) for Java methods. ASAP enhances the LLM’s prompt context by integrating static program analysis and information retrieval techniques to identify similar exemplar methods along with their developer-written Javadocs, and serves as the baseline in our study.

In contrast, we explore and compare the performance of four simpler approaches that do not require static program analysis, information retrieval, or the presence of exemplars as in the ASAP method. Our methods rely solely on the Java method body as input, making them lightweight and more suitable for rapid deployment in commercial software development environments.

We conducted experiments on an Ericsson software project and replicated the study using two widely-used open-source Java projects, Guava and Elasticsearch, to ensure the reliability of our results. Performance was measured across eight metrics that capture various aspects of similarity. Notably, one of our simpler approaches performed as well as or better than the ASAP method on both the Ericsson project and the open-source projects.

Additionally, we performed an ablation study to examine the impact of method names on Javadoc summary generation across our four proposed approaches and the ASAP method. By masking the method names and observing the generated summaries, we found that our approaches were statistically significantly less influenced by the absence of method names compared to the baseline. This suggests that our methods are more robust to variations in method names and may derive summaries more comprehensively from the method body than the ASAP approach.

\end{abstract}


\section{Introduction}
\label{sec:intro}
It is estimated nearly 60 to 80\% of software development costs are due to software maintenance. The chief difficulty of software maintenance arises due to developers not being able to understand or comprehend the underlying and legacy software code, which is often not originally written by them~\cite{giriprasad-icse11}. The burden of software development and program comprehension can be reduced through program comments and meaningful identifier names such as apposite (appropriate) method or function names~\cite{doc-for-maint,takang1996effects}. 

Comments can exist at various levels and can be of different types. A comment can be written before the method or within the body of a method. It typically describes the method's parameters, method's output, and the kind of exceptions that are thrown and handled by the method. It can explain the functionality of the piece of code and can be used as a reminder of tasks to be done (the TODO comments)~\cite{javadoc-guidelines}. Amongst the most useful comments are those that describe succinctly what a method does. Such summary comments are typically written before the method and appear in the API documentation of methods such as Javadoc.

The presence of well-written comments and meaningful identifier names positively affects the general commenting practices within the overall project. Although, developers are likely to add more comments in a project with well-written comments than in a project without comments~\cite{marin2005motivates}, in reality, developers under severe time constraints, limit adding good comments or updating existing comments. Thus, an automated system to generate comments tends to be helpful.

The advent of Generative AI and LLMs has impacted many areas, including software engineering practices \cite{llm-in-se-my-paper-cite, sridhara2023chatgpt, fan2023large} (e.g., code completion and comment generation).

We report our experiments, results and insights while automatically generating summary comments for Java methods within Ericsson\footnote{www.ericsson.com}, a global telecom vendor. We baseline the results against the recent LLM-based state-of-the-art (SOTA)  approach for code summarization - Automatic Semantic Augmentation of Prompts (ASAP) \cite{asap}. It uses static program analysis to extract data flow and other information from the body of a method and augments the LLM prompt. The notion of Chain-of-Thought prompting or few-shot learning are also leveraged to present sample input methods and their existing developer written summary comments to the LLM through prompts so that it can \emph{learn} the task of summary comment generation. The exemplar methods are not chosen in an ad-hoc manner but are those that are most similar to the input method for which the summary is desired to be generated (refer to Figure \ref{fig:asap}). The ASAP approach reports effectiveness on open source projects (including programming languages such as Python and Java).

In this paper, we have considered a \emph{closed source} Java project internal to the company and is specific to \emph{telecom} domain. Here, ASAP approach serves as baseline for benchmarking automatically generated summary comments for closed source telecom domain based Java methods. 

One of the challenges with our internal commercial software projects is that they have relatively less code duplication or cloning; hence retrieving similar existing methods to leverage as samples to the LLM is more challenging when compared with open source projects. This is particularly true for a newly written method which may need to be summarized. Thus, we wanted to explore alternate strategies that did not need additional samples.

Hence, we also explore the effectiveness of simpler prompting strategies in generating good Javadoc summary comments. We propose four different simpler prompting strategies that do not have the \emph{overhead} of the ASAP approach; \textit{viz.}, our approaches do not require static program analysis, do not require the presence of exemplar similar methods with existing developer written summaries and thus do not employ information retrieval (refer Figure \ref{fig:our-approaches}).

As an ablation study, we investigate the role of the method name in the generation of a summary comment. For example, consider the Java API List class method, \emph{add} whose Javadoc summary is "adds an item to the list". We wanted to explore if eliding or masking the method name adversely impacts the accuracy of the generated comment (across any approach). 

Somewhat \emph{surprisingly}, we found that simpler prompting approaches worked as well or better than ASAP in Java method summary comment generation task (across a wide variety of {\color{black}{eight}} different metrics). We also observe that this result holds with two open source projects, Guava and ElasticSearch and across two chosen LLMs as well (detailed in Section \ref{sec:eval})

Somewhat \emph{unsurprisingly}, we found that the actual method name (such as \emph{add}) played a role in the java doc summary comment generation.  However, results show that our approach is less susceptible in a statistically significant way to method name masking than the baseline ASAP approach. This indicates that our approach is more robust to method name choices and possibly derives the summary from the body of the method. We detail these observations through various metrics and significance tests in Section \ref{sec:eval}. 

To reiterate, the main contributions of this paper are as follows:
\begin{itemize}
    \item An experience report of using the LLM based ASAP method (SOTA approach) for code summarization of Java methods used within Ericsson. 
    \item A comparison of the above approach with other {\color{black}{four}} approaches proposed in our work. Our approaches involve prompt and context variations provided to the LLMs.
    \item The above studies are repeated using two LLMs: Llama and CodeLlama.
    \item For reproducibility, we also evaluate our and the ASAP approach on two popular large open source Java projects, \textit{viz.}, Guava and Elasticsearch.
    \item In an ablation study, we mask or elide the method name and check the performance of these approaches towards code summarization.
 \
\end{itemize}

The remainder of this paper is organized as follows: In Section~\ref{sec:approach} we describe our approach.
Section~\ref{sec:eval} delineates our evaluation, while in Section~\ref{sec:rel-work} we depict the related work and we conclude with a focus on future work in Section~\ref{sec:conc}.

\section{Approach}
\label{sec:approach}
In this section, we describe our approach towards automatic generation of summary comments for Java methods in Ericsson. We describe the baseline ASAP approach followed by approaches proposed in this work.
\subsection{The baseline approach: Automatic Semantic Augmentation of Prompts, ASAP}
\begin{figure*}[t]
\centering
\begin{subfigure}[t]{0.46\textwidth}
\includegraphics[width=0.99\columnwidth]{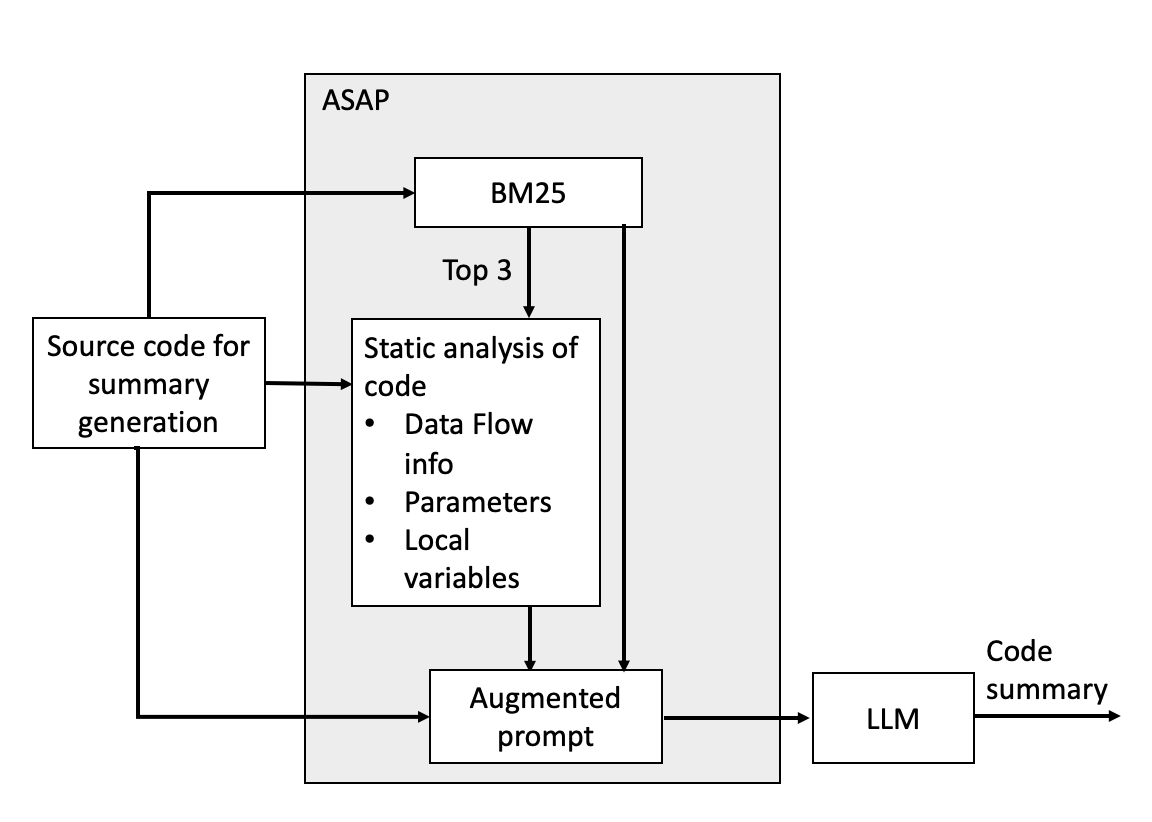}
\caption{The ASAP approach, figure adapted from ~\cite{asap}}
\label{fig:asap}
\end{subfigure}
~
\begin{subfigure}[t]{0.46\textwidth}
\includegraphics[width=0.99\columnwidth]{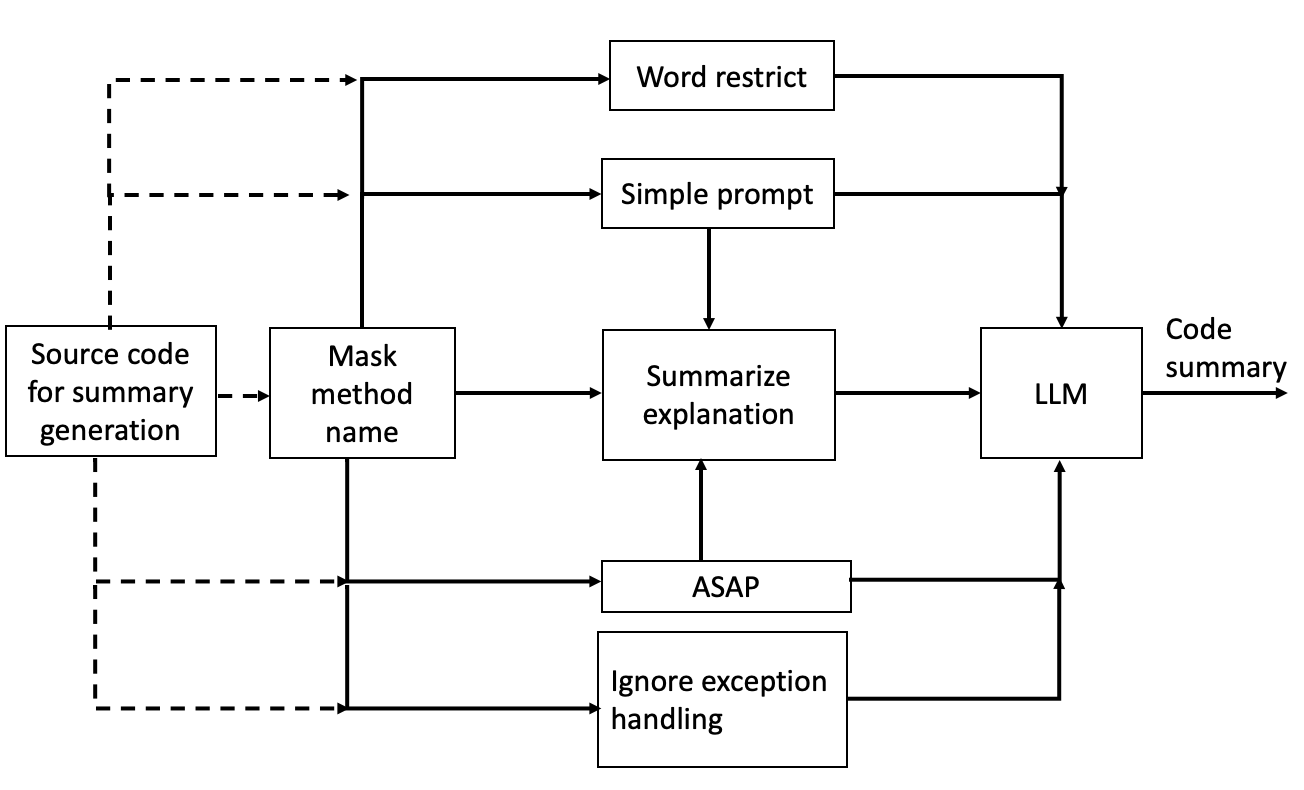}
\caption{Our Overall Study: Four approaches + ASAP}
\label{fig:our-approaches}
\end{subfigure}
\caption{Overview of the ASAP Approach and Our Study (prompts and contexts are shown in Table \ref{tab:approaches-javadoc-summary}).}
\end{figure*}

The ASAP approach is a black-box approach w.r.t. the use of LLMs. It does not modify the weights or the output probabilities of tokens by guiding the decoding stage as in ~\cite{agrawal2023monitor}. It augments the prompt using the following approaches:
\begin{enumerate}
    \item Semantic information from the source code of the method that needs to be summarized is introduced in prompt. In particular, data flow information of the variables in the method (so called def-use) is obtained using static analysis of the source code and introduced in prompt. In addition, identifiers in the method body such as parameters and local variables are also obtained and introduced in the prompt. 
    \item It presents three exemplars or samples of methods along with their existing summary comment. BM25 \cite{robertson1994some} is used to retrieve those methods that are most similar to the input method. For each of these three methods, the program analysis information as described in the first point are also included.
\end{enumerate}
Armed with the above additions to the prompt, the ASAP approach formulates the problem as a completion task to the LLM, i.e., complete the summary comment for the given method, given a sample of three methods with their existing summary and an input method, with all methods having the program analysis information such as data flow. The ASAP approach is shown in Figure~\ref{fig:asap}, interested readers can refer to ~\cite{ahmed2023improving, asap} for more details where it is described eloquently. 

In the following subsections we introduce our prompt-based approaches (in ascending order of "complexity") with the approach name introduced in heading of each subsection, depicted in Figure~\ref{fig:our-approaches} and summarized in Table \ref{tab:approaches-javadoc-summary}. 

\subsection{Simple Prompt} We explore (perhaps) the most simplest prompt i.e., "Please generate a Javadoc summary comment for the given Java method." and followed by the input method alone (i.e., method signature and body). With this, we empirically observed that the LLM would generate a verbose, almost verbatim, explanation of each statement (line) in the method.

\subsection{Simple Prompt with Word Length Restriction} Based on observation of previous prompt, we augmented the simple prompt with a request to the LLM to be concise. We hoped that asking it to restrict to twenty words would help in achieving brevity. Thus, this prompt was "Please generate a Javadoc summary comment for the given java method. Please use a maximum of twenty words for the summary." The length of twenty words was decided based on an observation of the median and maximum number of words in the Javadoc summaries in our project.

\subsection{Prompt to summarize the verbatim explanation given by the simple prompt} In this prompt, we provide the detailed explanation provided by the simple prompt, as the input (instead of the Java method). We then prompt the LLM to "Given the following JAVADOC indicated by JAVADOC: create a shorter summarized version of the JAVADOC. Keep the summary to under 20 words".

\subsection{Ignore Exception} This approach is inspired by ~\cite{giriprasad-summarization-ase10} and follows the Orwellian notion of "all are equal but some are more equal than others". In any Java method, all statements are equally important towards achieving the computational intent of the method. However, the language syntax and semantics dictate how the actual code is written and leads to many extraneous statements which are not needed for a succinct summary of the method. Consider for example, a method whose computational intent is to read all lines of a file. The Java syntax and semantics demands that the developer handles various exceptions, such as a FileNotFoundException, IOException and so on in a try-catch block. Such exception handling statements, while being absolutely imperative, add to the clutter and can obfuscate the intent of the method. Thus, such statements should be ignored from a summarization perspective. In addition to exception handling statements, logging statements and resource closing statements (the finally block in Java) can also be ignored for a succinct summary.

In this prompt, we request the LLM to ignore exception handling as - ``Please generate a Javadoc summary comment for the given Java method. While generating summary, please ignore: 1. exception handling (e.g. catch block); 2. resource cleanup (e.g. finally block); and, 3. logging statements (e.g. log).'' We describe our evaluation in the next section.

\begin{table}
    \centering
    \scalebox{0.85}{
    \begin{tabular}{|c|c|c|} \hline 
         \textbf{Approach Name}&  \textbf{Prompt}&  \textbf{Context} \\ \hline 
         Simple&  Please generate a Javadoc summary comment & Method body \\
         & for the Java method below &   \\ \hline 
         Word Restrict &   Please generate a Javadoc summary comment & Method body \\
         & for the Java method below.& \\
         & Please do not use more than 20 words."  & \\ \hline 
         SummarizeExplanation&  Given the following JAVADOC indicated by JAVADOC: & The verbatim Natural Language \\
         & create a shorter summarized version of the JAVADOC. & explanation of the java method,  \\
         & Keep the summary under 20 words. & produced by the simple prompt.  \\
         \hline 
         IgnoreExceptionHandling&  
         Please generate a Javadoc summary comment & Method body \\
         & for the Java method below. & \\
         & While generating summary, please ignore: & \\
         & 1. exception handling (e.g. catch block) & \\
         & 2. resource cleanup (e.g. finally block) & \\  & 3. logging statements (e.g. log) & \\
         &  & \\ \hline 
         ASAP~\cite{asap} & Write down the original comment & Method body with 3 similar methods along \\
         & written by the developer.& with developer comments. Data Flow \\
         & Comment: & and Identifier information \\ 
         & & of all methods. \\
         \hline
    \end{tabular}
    }
    \caption{Approaches to Generate JavaDoc Summary}
    \label{tab:approaches-javadoc-summary}
\end{table}

\section{Evaluation}
\label{sec:eval}
\subsection{Experimental Subjects}
We consider a large Java project within our Ericsson code base and also two popular open source Java projects, \textit{viz.}, Guava~\cite{guava}
and ElasticSearch~\cite{elasticsearch}. The statistics about the projects such as number of methods and methods with a leading Javadoc comment are shown in Table~\ref{tab:project-loc-etal-stats}.


\subsection{Large Language Models Used}
\label{sec:llm}
In this section, we discuss the rationale for the LLMs used in our study. We explored several models to arrive at our final chosen set of LLMs. Our chief criteria was based on the following two factors:
\begin{enumerate}
    \item The LLM should have a commercial friendly license.
    \item We should be able to host the LLM within our server network.
\end{enumerate}

This ruled out popular LLMs such as OpenAI\footnote{https://openai.com/policies/terms-of-use} and thus, we chose the Llama-70B (referred to as Llama in rest of the paper) and CodeLlama from Llama family (by Meta/Facebook) and the Mixtral LLM (from Mistal AI)~\cite{rozière2023code,touvron2023llama,mixtral}. The LLMs are hosted on a server with one Tesla A100 GPU and 1 TB RAM along with an AMD EPYC 74F3 24-Core Processor CPU.

\subsection{Research Questions}
\textbf{RQ1:} How does the SOTA baseline approach, ASAP, perform in generating summary comments for a commercial telecom domain project? Alternately, what is the best prompt and context strategy? Do simpler approaches work as well or better than ASAP?

\textbf{RQ2:} What is the role of the method name in the generation of the method's summary? Does eliding or masking the method name adversely impact the performance of the various approaches in generating the summary?

\begin{table}
    \centering
    \begin{tabular}{|c|c|c|c|c|} \hline 
         &  LOC&  Classes&  Methods & Methods \\
         &  &  &   & with Javadoc\\ \hline 
         Ericsson Project &  706,724&  6253&  51984& 12568\\ \hline
         Elasticsearch~\cite{elasticsearch} &  1,071,602 & 6623 & 68635 & 17562\\ \hline
         Guava~\cite{guava} & 177,843 & 604 & 11777 & 4321 \\\hline
    \end{tabular}
    \caption{Subject Details: LOC = Lines Of Code}
    \label{tab:project-loc-etal-stats}
\end{table}

\subsection{Methodology} We used the JavaParser API~\cite{javaparser} to parse the source code, build the Abstract Syntax Tree (AST) and extract methods with Javadoc. This information is captured in JSON format as expected by ASAP. We also used the source code provided by the ASAP to extract the identifiers and data flow graph and build the set of similar methods. We modified the ASAP code to generate the final prompt and context which we then used with our chosen set of LLMs.

\textbf{ASAP settings:} We use Data Flow Graph (DFG) information and identifier information. We don't use the Repo information. We find three similar sample methods (default) as proposed in \cite{asap}. We use ASAP's code to obtain the above and also the final prompt that they could have used with OpenAI LLM (as per \cite{asap}).

We divided the Java methods with Javadoc into two sets of train and evaluation in a 80:20 manner as desired by ASAP (Our approaches \emph{do not} require any training data.). We did the following to ensure that we have non-trivial evaluation methods.
\begin{enumerate}
    \item We filtered out methods with less than ten lines of code as many of these are simple get/set kind of methods.
    \item We then gathered 100 random methods from the test set and sorted them in descending order of lines of code. We did this to ensure that we had fairly long (not too small) methods.
\end{enumerate}

Note that the ASAP finds similar methods to the input as exemplars. For this, it relies on splitting the code into tokens. We used the same tokenization strategy as used by ASAP, although other tokenization strategies are possible. 

\textbf{Masking approach:} 
For the masking experiments to answer \emph{RQ2}, we simply replace the method name with the word `MASKED'. Thus, for example, in Listing~\ref{lst:remove}, the method name \emph{remove} is replaced by \emph{MASKED}, with everything else remaining the same.

\textbf{Evaluation Metrics Computation:} We use the metrics available in ASAP code \cite{asap} to evaluate performance of ASAP and our approaches. The ASAP work reports BLEU-CN \cite{alon2018code2seq} (leverages modified Laplacian smoothing), BLEU-DC \cite{hu2020deep} (leveraging another smoothing method), METEOR \cite{banerjee2005meteor} and ROUGE-L \cite{lin2004rouge} metrics; note that all of these are some modified n-gram techniques proposed for text summarization and translation tasks.  

In this work, we additionally also introduce BERTScore \cite{zhang2019bertscore}, BLEU-RT \cite{sellam2020bleurt} and Sentence Similarity (based on cosine similarity using miniLM \cite{wang2020minilm} for embedding) metrics to score semantic similarity aspects of generated summary with that of ground truth. The wide variety of metrics capture different aspects of similarity between the generated summary comment and the developer's original comments, which serves as the ground truth. Note that the implementation of BLEU-CN as per \cite{asap} is \emph{slightly different} than using the BLEU-CN library package - for a \emph{fair} comparison, we use the ASAP's implementation and refer it to as BLEU in the plots and tables. 

We report the distribution of the metrics (using mean, standard deviation) in Tables \ref{tab:meanStdCommercialCodeLLama} and  
\ref{tab:meanStdCommercialLLAMA2}
for CodeLlama, Llama, respectively (using commercial datasets) and for selected prompts on open source datasets in Table \ref{tab:meanStdOpenSource} on Codellama, Llama and Mixtral LLMs. 

In addition, we report distribution of various metrics of best performing prompt against the baseline along with the p-values for one-sided t-test and KS-tests; these metrics are also reported post masking the method name of the source code while introducing the query in the prompt (Figures \ref{fig:comparisonMetricscodellama}, \ref{fig:comparisonMetricsllama70b}, \ref{fig:maskingEffect}). The distribution of best prompt across the selected 100 queries is reported in Figure \ref{fig:bestPrompt}.

\subsection{Results}

We present here experimental results of our work. We discuss the responses and their evaluation for \emph{RQ1} and \emph{RQ2}.

We evaluate performance of our simpler approaches  against the ASAP approach. We have five approaches; each approach has two variants with the method name masked or unmasked in the input, leading to results on ten approaches. Evaluation involves comparing values based on eight different metrics.

For each of the 100 methods in our commercial evaluation set, we generate the summary for each approach, compute all metrics' scores, aggregate and report the mean and standard deviation of the metric scores (and median). The mean (and standard deviations) are shown in  Table~\ref{tab:meanStdCommercialCodeLLama} and Table~\ref{tab:meanStdCommercialLLAMA2} for CodeLlama and Llama LLMs, respectively.

From the tables, it can be seen that with the CodeLlama LLM Table~\ref{tab:meanStdCommercialCodeLLama}, our approach reports a slightly higher mean in seven of eight metrics. With the Llama LLM (Table~\ref{tab:meanStdCommercialLLAMA2}), our approach had a higher mean value on \emph{all} metrics.

Note that we report the mean and not the median, because the ASAP paper~\cite{asap} reports the mean. However, the comparison of the median scores were even more favorable to our approaches. The median scores of any of our approaches was better than the median scores of ASAP on all 8 metrics with CodeLlama and on 7 out of 8 metrics with Llama-70B. (not shown in any Table to avoid clutter)~\footnote{There was a tie between ASAP and our approach on \emph{rouge-recall} metric.}. 

\textbf{Conclusion:} These results show that our simpler approaches are able to perform as well or better than ASAP for the task of summary comment generation in a commercial Java project on methods which have  more than 10 lines of code.

\subsubsection{\textbf{Generalizability of results:}}
To check generalizability of the previous results for proposed approaches, we repeat the experiments on two open source projects, Guava and Elasticsearch. We used only the best of our approaches, \emph{wordrestrict prompt} to compare with ASAP. Table~\ref{tab:meanStdOpenSource} shows the mean and standard deviation across the eight metrics across three LLMs. Note that, we used the Mixtral LLM here in addition to Llama and CodeLlama.

Again, the trend of our approach having a higher mean (with lower standard deviation) and higher median repeats across a \emph{majority} of the metrics and across \emph{all} three LLMs.

\subsubsection{\textbf{Statistical Significance:}}

The Table \ref{tab:meanStdCommercialCodeLLama} and Table\ref{tab:meanStdCommercialLLAMA2} clearly indicate that among the 4 proposed prompts, the \emph{wordrestrictprompt} shows improved performance over the baseline, ASAP. To validate this, we ran a set of statistical tests for our best performing prompt against the baseline. In particular, for each of the metrics we ran a one-sided t-test with the null hypothesis that the means of the two distributions (\textit{viz.} our best prompt and baseline) are equal and the alternate hypothesis that our means is higher than the means of the baseline. We also repoert one-sided Kolmogorov–Smirnov test outcomes \cite{berger2014kolmogorov} (KS test) with the null hypothesis that the metrics for our best prompt and baseline are drawn from the same distribution and the alternate hypothesis that our metrics are stochastically greater than that of the baseline.

\begin{figure*}[htbp]
    \centering
    \begin{subfigure}[b]{0.75\textwidth}
        \centering
        \includegraphics[width=\textwidth]{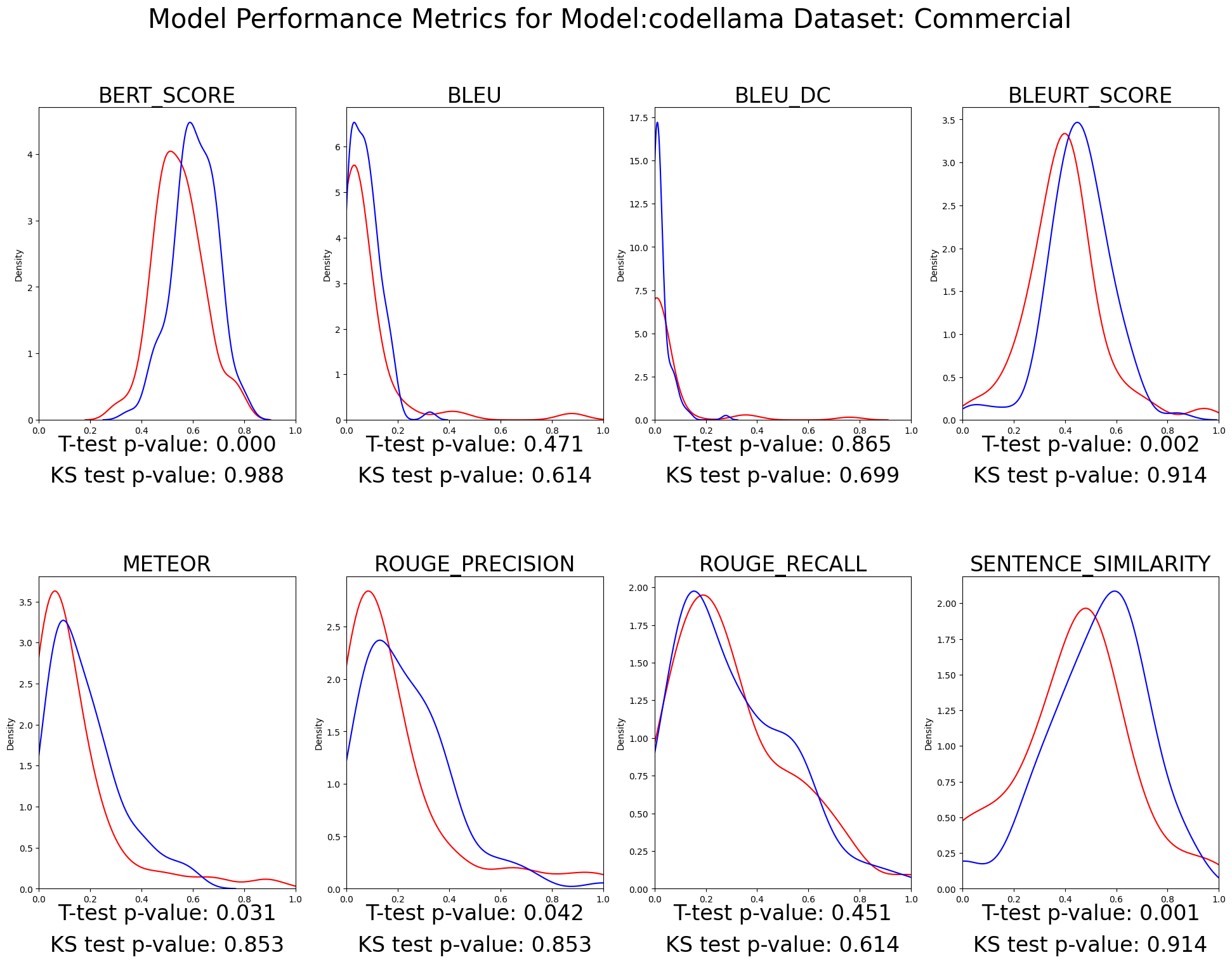} 
        \caption{Codellama on Commercial Project} 
        \label{fig:comparisonMetricscodellama} 
    \end{subfigure}
    
    \begin{subfigure}[b]{0.75\textwidth}
        \centering
        \includegraphics[width=\textwidth]{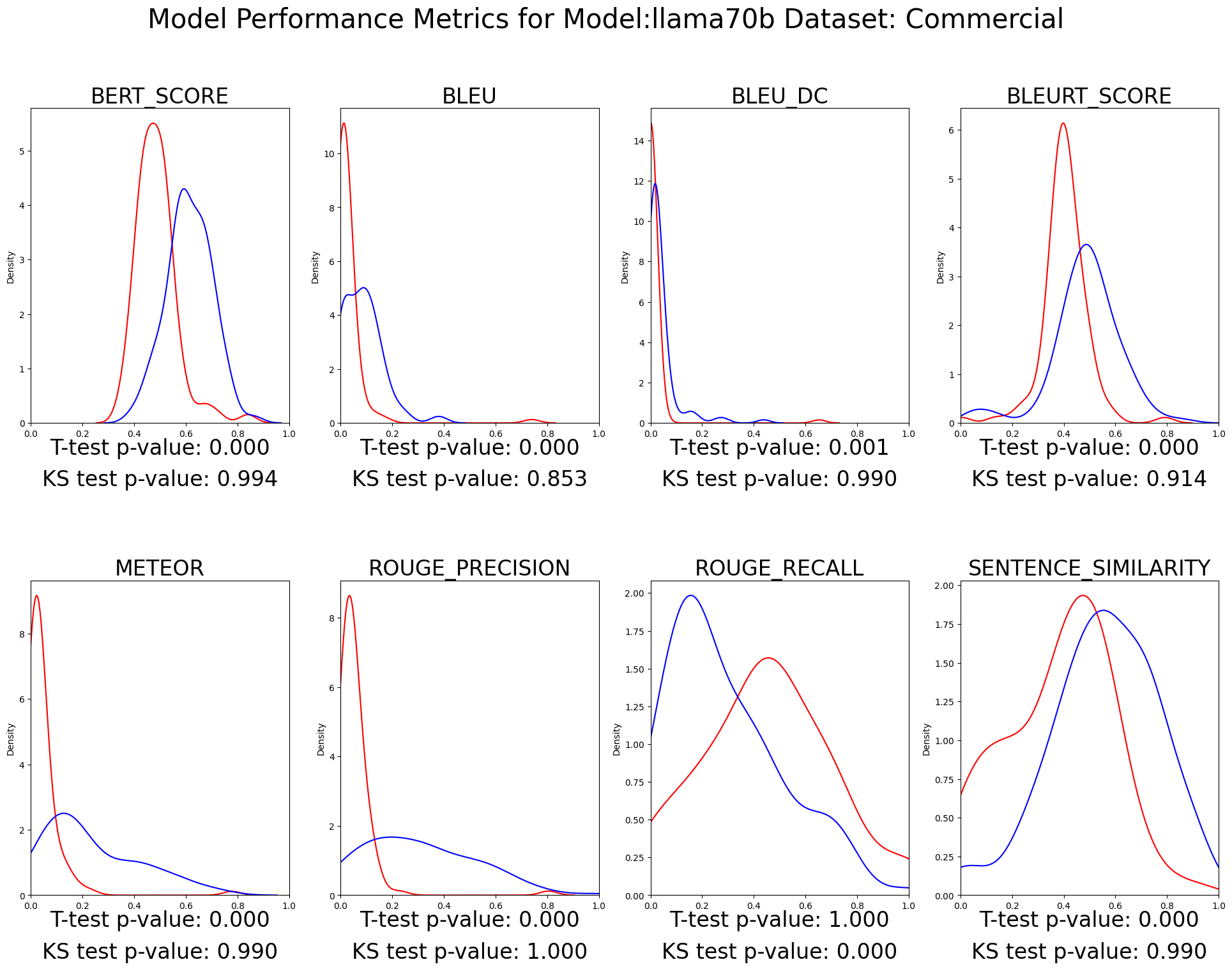} 
        \caption{Llama70b on Commercial Project} 
        \label{fig:comparisonMetricsllama70b} 
    \end{subfigure}
    \caption{Distribution of various metrics of our best prompt \textbf{wordrestrict} in {\color{blue} blue} with baseline prompt in {\color{red} red}. We show the distributions and p-values for one-sided t-test for means and one-sided KS test. The alternative hypothesis for the t-test is that our prompt has a higher mean than baseline, whereas the alternate hypothesis for the KS test is that our prompt is stochastically greater than the baseline. Metrics having p-values $<0.05$ can be considered as statistically significant.}
    \label{fig:comparisonMetrics}
\end{figure*}

Our results shown below each metric in Fig. \ref{fig:comparisonMetrics} show that for CodeLlama (sub-figure \ref{fig:comparisonMetricscodellama}), our  mean is higher than the baseline in 5 of 8 metrics but not stochastically greater than the baseline in any of metrics. Similarly, with Llama (sub-figure \ref{fig:comparisonMetricsllama70b}), our best prompt mean is higher than the baseline in 7 of the 8 metrics and stochastically greater in one of them.  The results validate that the proposed prompt performs better than the baseline, especially on average, across multiple metrics.  

\begin{figure*}[htbp]
    \centering
    \begin{subfigure}[b]{0.75\textwidth}
        \centering
        \includegraphics[width=\textwidth]{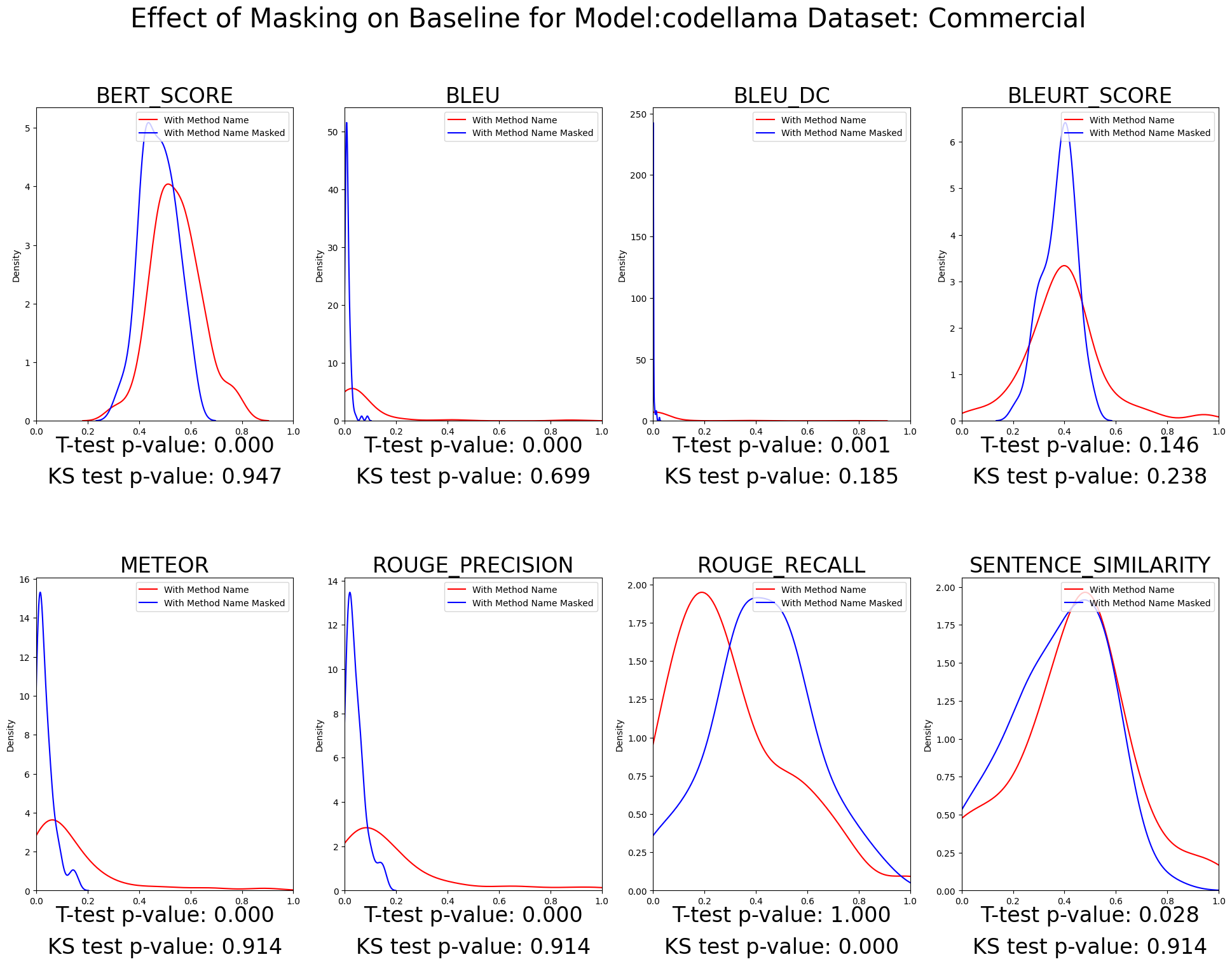} 
        \caption{Masking Effect on Baseline Prompt} 
    \label{subfig:baselinemasking}
    \end{subfigure}
    
    \begin{subfigure}[b]{0.75\textwidth}
        \centering
        \includegraphics[width=\textwidth]{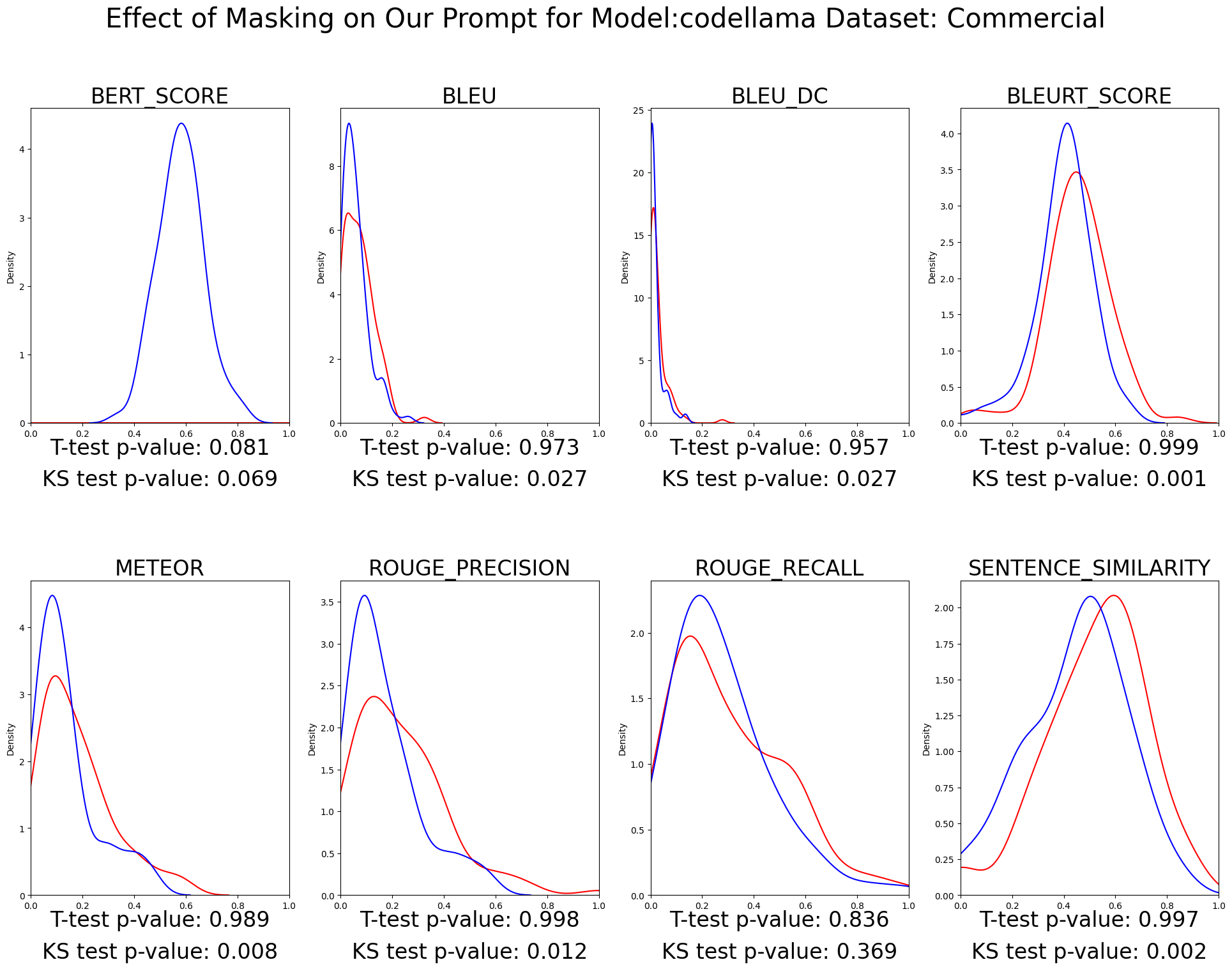} 
        \caption{Masking Effect on our Prompt} 
        \label{fig:MaskingOurs} 
    \end{subfigure}
    \caption{Effect on method name masking on distribution of various metrics for baseline prompt on the left  and our best prompt \textbf{wordrestrict} on the right   evaluated on the commercial dataset. The {\color{blue} blue} distributions indicate prompt with the method name available, and the {\color{red} red} distributions indicate prompt with the method name masked. We show the distributions and p-values for one-sided t-test for means and one-sided KS test. The alternate hypothesis for the t-test is that the metric has a higher mean with method name unmasked than method name masked whereas the alternative hypothesis for the KS-test is that the metric with method name unmasked is stochastically greater than with method name masked. Metrics having p-values $<0.05$ can be considered as statistically significant.}
    \label{fig:maskingEffect}
\end{figure*}

We run similar statistical tests for understanding the robustness of the approaches to method name masking. For the baseline ASAP and our best prompt (\emph{wordrestrictprompt}), we separately compare means and distributions with the method name available and masked approach for each of the 8 metrics and LLM models. For brevity, we show the results only for CodeLlama in Fig. \ref{fig:maskingEffect}. We run a one-sided t-test, separately for baseline and our best prompt, with the null hypothesis that means for masked and unmasked are equal and the alternate hypothesis that the means for unmasked are higher than that of the method with the name masked. Similarly, we run a KS-test for each metric, independently for ASAP and \emph{wordrestrictprompt}, with the null hypothesis that the distribution with method name unmasked is stochastically greater than the distribution with masked. 

Our results (Fig. \ref{fig:maskingEffect}), show that our prompt is less sensitive to method name choice. In particular, for the baseline prompt shown in sub-figure \ref{subfig:baselinemasking} in 6 (out of 8 metrics) of the  means with method name available (i.e. unmasked) are greater than those with the method names masked. On the other hand,  for none of the metrics using our prompt the means with unmasked method name is higher than masked, in a statistically significant sense. This shows that our method is more robust to method name choices and gives an indication of the fact that the summary produced by our approach is derived in a more complete manner from the method body itself.


\begin{figure*}[htbp]
        \centering
        \includegraphics[width=\textwidth]{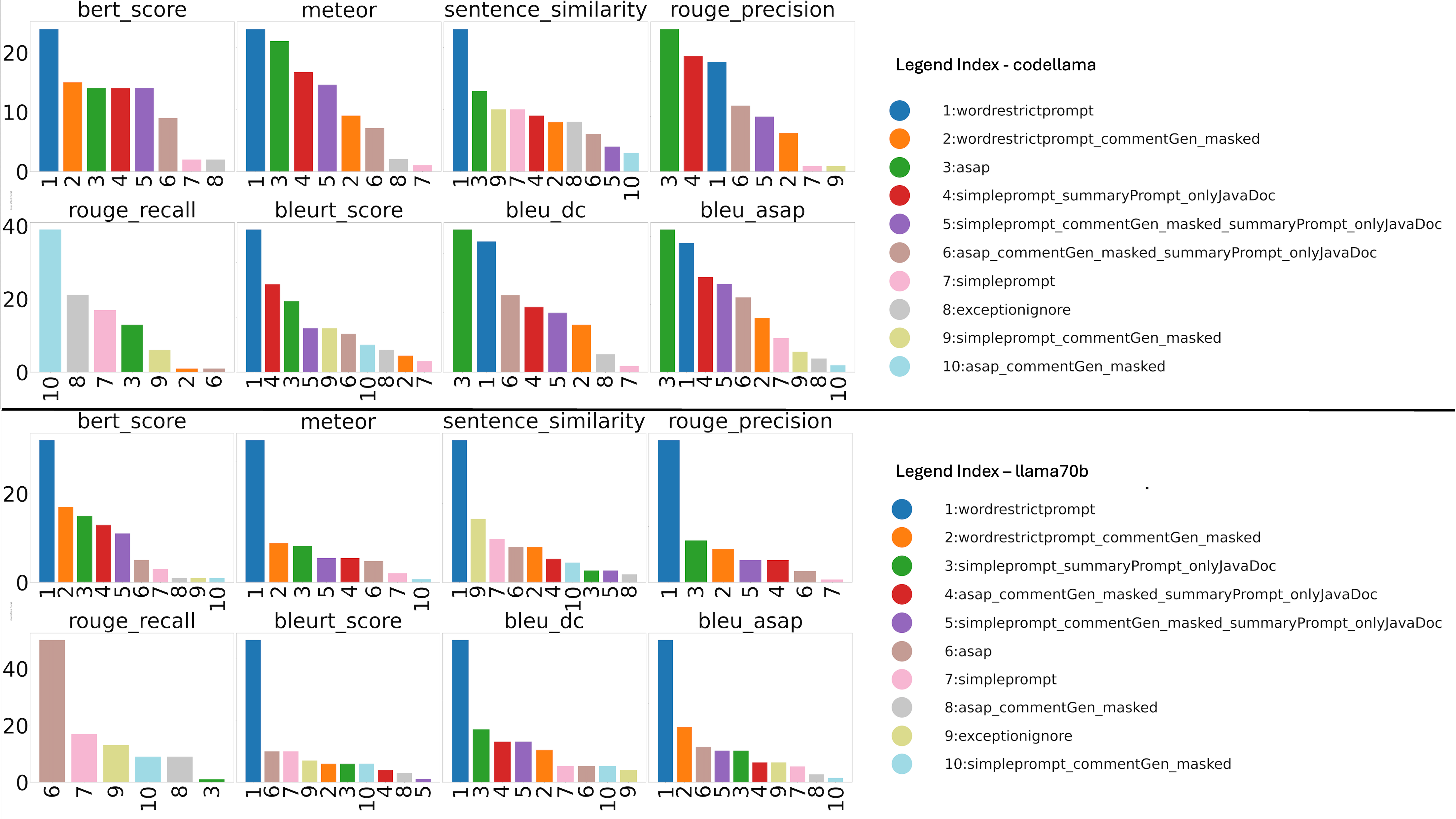} 
       \caption{Distribution of the best prompt (approach) across 100 queries (methods) on commercial dataset. There are eight charts corresponding to eight metrics in each sub-figure. In each chart, the vertical bars denote how many times a certain prompt scored best on a certain metric. The metric name is on top of the charts. Our approach leads in 4 of 8 metrics with CodeLlama (top Sub-figure). Our approach leads in 7 of 8 metrics with Llama-70b (bottom sub-figure). }
  
    \label{fig:bestPrompt}
\end{figure*}

\begin{table}[h]
\centering
\scalebox{0.7}{\begin{tabular}{|c|c|c|c|c|c|c|c|c|c|}
\hline
    \textbf{prompt} & \textbf{bert\_score} & \textbf{bleu\_dc} & \textbf{bleu} & \textbf{bleu-rt} & \textbf{meteor} & \textbf{rouge\_prec} & \textbf{rouge\_rec} & \textbf{sent\_sim} \\ \hline

asap                                            & 0.55(0.1)                                  & \textbf{0.04(0.12)} & 0.07(0.14)                              & 0.4(0.15)                            & 0.14(0.18)                          & 0.18(0.22)                               & 0.3(0.23)                               & 0.43(0.23)                             \\
asap\_commentGen\_masked                        & 0.48(0.07)                                 & 0(0)                                       & 0.01(0.01)                              & 0.38(0.06)                           & 0.04(0.03)                          & 0.04(0.04)                               & 0.42(0.2)                               & 0.37(0.19)                             \\
asap\_commentGen\_masked\_summaryPrompt         & 0.57(0.07)                                 & 0.02(0.03)                                 & 0.06(0.05)                              & 0.38(0.13)                           & 0.13(0.11)                          & 0.17(0.14)                               & 0.21(0.18)                              & 0.36(0.21)                             \\
exceptionignore                                 & 0.51(0.08)                                 & 0(0.01)                                    & 0.03(0.02)                              & 0.41(0.06)                           & 0.07(0.07)                          & 0.08(0.07)                               & 0.42(0.22)                              & 0.49(0.18)                             \\
simpleprompt                                    & 0.52(0.09)                                 & 0.01(0.02)                                 & 0.03(0.03)                              & 0.41(0.08)                           & 0.07(0.07)                          & 0.09(0.07)                               & \textbf{0.44(0.24)}                     & 0.51(0.18)                             \\
simpleprompt\_commentGen\_masked                & 0.51(0.08)                                 & 0(0.01)                                    & 0.02(0.02)                              & 0.4(0.07)                            & 0.06(0.05)                          & 0.07(0.06)                               & 0.41(0.21)                              & 0.46(0.17)                             \\
simpleprompt\_commentGen\_masked\_summaryPrompt & 0.58(0.09)                                 & 0.02(0.02)                                 & 0.06(0.04)                              & 0.4(0.12)                            & 0.14(0.1)                           & 0.18(0.11)                               & 0.26(0.18)                              & 0.43(0.2)                              \\
simpleprompt\_summaryPrompt                     & 0.59(0.09)                                 & 0.02(0.03)                                 & \textbf{0.07(0.05)}                     & 0.44(0.13)                           & 0.18(0.13)                          & \textbf{0.24(0.18)}                      & 0.27(0.2)                               & 0.5(0.2)                               \\
wordrestrictprompt                              & \textbf{0.61(0.08)} & 0.03(0.04)                                 & 0.07(0.06)                              & \textbf{0.45(0.13)}                  & \textbf{0.18(0.14)}                 & 0.23(0.18)                               & 0.3(0.22)                               & \textbf{0.52(0.19)}                    \\
wordrestrictprompt\_commentGen\_masked          & 0.58(0.09)                                 & 0.02(0.03)                                 & 0.06(0.05)                              & 0.4(0.11)                            & 0.14(0.12)                          & 0.17(0.14)                               & 0.28(0.19)                              & 0.45(0.19) \\
\hline
\end{tabular}}
\caption{Means (and standard deviations) for best different metrics and prompts on commercial project using Codellama as LLM. Best score for each metric is in BOLD.
In case of ties in mean, score with lower standard deviation is chosen as best.}
\label{tab:meanStdCommercialCodeLLama}
\end{table}

\begin{table}[htbp]
\centering
\scalebox{0.75}{\begin{tabular}{|c|c|c|c|c|c|c|c|c|c|}
\hline
    \textbf{prompt} & \textbf{bert\_score} & \textbf{bleu\_dc} & \textbf{bleu} & \textbf{bleu-rt} & \textbf{meteor} & \textbf{rouge\_prec} & \textbf{rouge\_rec} & \textbf{sent\_sim} \\ \hline
asap                                            & 0.49(0.08)                               & 0.01(0.07)                            & 0.03(0.08)                              & 0.4(0.09)                            & 0.04(0.08)                          & 0.05(0.08)                               & 0.45(0.25)                              & 0.38(0.21)                             \\
asap\_commentGen\_masked                        & 0.48(0.07)                               & 0(0.01)                               & 0.02(0.04)                              & 0.4(0.08)                            & 0.04(0.05)                          & 0.05(0.05)                               & 0.43(0.25)                              & 0.35(0.19)                             \\
asap\_commentGen\_masked\_summaryPrompt         & 0.57(0.07)                               & 0.01(0.01)                            & 0.05(0.04)                              & 0.35(0.11)                           & 0.11(0.1)                           & 0.14(0.14)                               & 0.22(0.19)                              & 0.32(0.19)                             \\
exceptionignore                                 & 0.52(0.08)                               & 0(0.02)                               & 0.03(0.03)                              & 0.42(0.06)                           & 0.07(0.05)                          & 0.08(0.06)                               & 0.44(0.22)                              & 0.52(0.18)                             \\
simpleprompt                                    & 0.52(0.08)                               & 0(0.01)                               & 0.03(0.03)                              & 0.44(0.05)                           & 0.07(0.06)                          & 0.09(0.06)                               & \textbf{0.46(0.24)}                     & 0.52(0.18)                             \\
simpleprompt\_commentGen\_masked                & 0.5(0.08)                                & 0(0)                                  & 0.02(0.02)                              & 0.42(0.06)                           & 0.06(0.04)                          & 0.07(0.05)                               & 0.43(0.21)                              & 0.45(0.18)                             \\
simpleprompt\_commentGen\_masked\_summaryPrompt & 0.57(0.09)                               & 0.02(0.02)                            & 0.05(0.04)                              & 0.37(0.1)                            & 0.12(0.09)                          & 0.15(0.1)                                & 0.23(0.17)                              & 0.4(0.17)                              \\
simpleprompt\_summaryPrompt                     & 0.58(0.1)                                & 0.02(0.03)                            & 0.06(0.05)                              & 0.41(0.1)                            & 0.17(0.13)                          & 0.2(0.15)                                & 0.28(0.2)                               & 0.47(0.18)                             \\
wordrestrictprompt                              & \textbf{0.61(0.09)}                      & \textbf{0.04(0.06)}                   & \textbf{0.09(0.08)}                     & \textbf{0.49(0.13)}                  & \textbf{0.24(0.18)}                 & \textbf{0.31(0.21)}                      & 0.29(0.22)                              & \textbf{0.56(0.2)}                     \\
wordrestrictprompt\_commentGen\_masked          & 0.6(0.09)                                & 0.02(0.04)                            & 0.07(0.07)                              & 0.42(0.13)                           & 0.16(0.14)                          & 0.2(0.17)                                & 0.22(0.17)                              & 0.45(0.22)  \\
\hline
\end{tabular}}
\caption{Means (and standard deviations) for best different metrics and prompts on commercial project using llama2-70b as LLM. Best score for each metric is in 
BOLD.
In case of ties in mean, score with lower standard deviation is chosen as best.}
\label{tab:meanStdCommercialLLAMA2}
\end{table}

\begin{table}[ht]
    \scalebox{0.75}{\begin{tabular}{|c|c|c|c|c|c|c|c|c|c|c|}
        \hline
    \textbf{Dataset} &   \textbf{prompt} & \textbf{LLM} & \textbf{bert\_score} & \textbf{bleu\_dc} & \textbf{bleu} & \textbf{bleu-rt} & \textbf{meteor} & \textbf{rouge\_prec} & \textbf{rouge\_rec} & \textbf{sent\_sim} \\ \hline

    guava               & asap               & codellama & 0.57(0.06)                      & 0.01(0.03)                   & 0.03(0.03)                     & 0.39(0.06)                  & 0.1(0.08)                  & 0.11(0.09)                      & \textbf{0.35(0.18)}            & 0.51(0.16)                    \\
                        & wordrestrictprompt & codellama & \textbf{0.63(0.08)}             & \textbf{0.05(0.11)}          & \textbf{0.06(0.11)}            & \textbf{0.47(0.09)}         & \textbf{0.23(0.14)}        & \textbf{0.38(0.2)}              & 0.19(0.17)                     & \textbf{0.55(0.15)}           \\
                        & asap               & llama70b  & 0.56(0.11)                      & 0.04(0.12)                   & \textbf{0.08(0.17)}            & 0.42(0.12)                  & 0.13(0.18)                 & 0.15(0.19)                      & \textbf{0.44(0.25)}            & 0.47(0.21)                    \\
                        & wordrestrictprompt & llama70b  & \textbf{0.63(0.08)}             & \textbf{0.04(0.05)}          & 0.04(0.07)                     & \textbf{0.47(0.09)}         & \textbf{0.25(0.13)}        & \textbf{0.46(0.19)}             & 0.17(0.15)                     & \textbf{0.57(0.15)}           \\
                        & asap               & mixtral   & \textbf{0.6(0.09)}              & \textbf{0.04(0.12)}          & \textbf{0.08(0.15)}            & 0.44(0.12)                  & 0.18(0.19)                 & 0.25(0.27)                      & 0.4(0.27)                      & 0.51(0.2)                     \\
                        & wordrestrictprompt & mixtral   & 0.57(0.06)                      & 0.02(0.06)                   & 0.04(0.07)                     & \textbf{0.45(0.09)}         & \textbf{0.18(0.12)}        & \textbf{0.37(0.21)}             & \textbf{0.16(0.17)}            & \textbf{0.55(0.13)}           \\ \hline
    elasticsearch & asap               & codellama & 0.5(0.07)                       & 0(0.01)                      & 0.01(0.01)                     & 0.38(0.06)                  & 0.05(0.05)                 & 0.05(0.05)                      & \textbf{0.41(0.19)}            & 0.4(0.18)                     \\
                        & wordrestrictprompt & codellama & \textbf{0.57(0.1)}              & \textbf{0.02(0.04)}          & \textbf{0.05(0.05)}            & \textbf{0.39(0.12)}         & \textbf{0.13(0.11)}        & \textbf{0.17(0.15)}             & 0.27(0.19)                     & \textbf{0.44(0.22)}           \\
                        & asap               & llama70b  & 0.49(0.06)                      & 0(0)                         & 0.01(0.01)                     & 0.39(0.08)                  & 0.03(0.04)                 & 0.04(0.04)                      & \textbf{0.41(0.21)}            & 0.36(0.19)                    \\
                        & wordrestrictprompt & llama70b  & \textbf{0.59(0.09)}             & \textbf{0.02(0.07)}          & \textbf{0.06(0.09)}            & \textbf{0.43(0.11)}         & \textbf{0.16(0.15)}        & \textbf{0.23(0.21)}             & 0.27(0.2)                      & \textbf{0.48(0.18)}           \\
                        & asap               & mixtral   & 0.55(0.07)                      & 0.01(0.01)                   & \textbf{0.04(0.04)}            & 0.37(0.11)                  & 0.11(0.1)                  & 0.16(0.16)                      & \textbf{0.27(0.2)}             & 0.42(0.19)                    \\
                        & wordrestrictprompt & mixtral   & \textbf{0.56(0.06)}             & \textbf{0.02(0.03)}          & 0.04(0.05)                     & \textbf{0.4(0.12)}          & \textbf{0.12(0.11)}        & \textbf{0.19(0.16)}             & 0.23(0.18)                     & \textbf{0.45(0.18)} \\ \hline        
    \end{tabular}}
    \caption{Means (and standard deviations) for best different metrics, prompts and LLMs on two open source datasets. We have shown results only for our best prompt (wordrestrictprompt)  from the commercial dataset and baseline. The best score for each metric for each dataset and model combination is in BOLD.
    In case of ties in mean, score with lower standard deviation is chosen as best.}
    \label{tab:meanStdOpenSource}
    \end{table}


\subsubsection{\textbf{Best Prompt analysis:}}
\textbf{Across the 100 evaluation methods in our commercial project, which approaches (prompts) performed the best?} Figure~\ref{fig:bestPrompt} shows the distribution of the \emph{best} prompt (approach) across the 100 methods we had in our evaluation set for the Ericsson commercial project. The top part shows the results with the CodeLlama, and the bottom part refers to results using Llama. 

There are eight bar charts apiece for Codellama and Llama. Each chart (of 8 bar charts within a subplot) corresponds to a metric that we have used (from the implementation in the ASAP approach~\cite{asap}). The metric name is on top of each chart. Within each chart, the bar chart depicts the distribution i.e., the number of times a particular approach had the highest score compared to other approaches.

We can see from the charts, that with CodeLlama and \emph{WordRestrict} prompt shows better values in four (of 8) metrics, with the ASAP approach in three (of 8) occasions. Even for these three metrics, our \emph{WordRestrict} prompt shows competitive performance. 

The Llama LLM in combination with \emph{WordRestrict} prompt shows higher values in seven of the eight metrics.

\textbf{Conclusion:} Considering the performance of each approach on each of the 100 commercial methods, we find that our prompt based approach, in particular, the \emph{WordRestrict} prompt performs as well or better than the ASAP, across two different LLMs and eight different comment (text) similarity metrics.


\subsection{Threats to Validity}
Our study is done on a commercial telecom Java project and thus may not generalize to open source projects and also to non-Java projects. To mitigate this, we repeated our study on two large open source Java projects, \textit{viz.}, Guava and Elasticsearch. In future, we will experiment with other languages used in Ericsson.

Note that ASAP approach is language agnostic in some ways as it works with any programming language amenable to static analysis of source code. Our approach is further lightweight and only requires the method or function body. Thus, our approach should be generalizable across languages.

We also used Meta's Llama and Codellama as the LLMs in our work based on criteria listed in Section \ref{sec:eval}. We do not use OpenAI and other API based LLMs as they are not commercial friendly and thus we are not certain of the generalizability of our approach across all popular LLMs.

To mitigate the above threat of lack of generalizability across LLMs, we checked the robustness of our techniques with a separate LLM altogether and we chose the recently released Mixtral 8*7B Mixture of Experts (MoE) model. Our results for these are shown in Table \ref{tab:meanStdOpenSource}. The results indicate robustness of our prompts across the chosen LLMs.
For few of the metrics, the baseline prompt does better on the Mixtral model - this could be due to the way this model (coming from a different organization) may expect it's prompt to be structured. This needs further investigation to understand and design techniques to make prompts robust to LLM change.

\subsection{Discussion}
We observe from Fig \ref{fig:bestPrompt} that the performance of \emph{WordRestrict} approach is better than that of baseline. The p-values shows that the improvement is statistically significant. From Figure \ref{fig:maskingEffect}, we see that the decrease in terms of metric values after masking the method name is significant for baseline approach in 6 out of 8 metrics and 4 out of 8 on our best prompt approach. This implies that much of the code summarization task is influenced by the method name than the logic or flow of the code itself. This is a noteworthy observation as this implies providing correct names to methods can aid in code summarization performance. Thus, method names play a crucial role in summary comment generation as evinced by the reduced metric scores when the method name is masked. This holds even with masking of method names in the ASAP approach as well. Note that ASAP prompts are augmented with Data Flow Graphs, Parameters and local variables as well. Our results (refer Figure \ref{fig:bestPrompt}) suggest that simpler prompts with some fine tuning such as requesting the LLM to be concise and prefer brevity does better than the ASAP approach. 


\subsection{Samples of Generated Summary Comments}
Table~\ref{tab:samples-gen-sum} juxtaposes summary comments written by the developer with the summaries generated by our approach (\emph{WordRestrict} Prompt) and the ASAP. Note that to maintain confidentiality we have obfuscated the class names and denote them as X, Y, Z etc.

Consider the example 1 in Table~\ref{tab:samples-gen-sum}.
Our approach correctly identifies the operation (\textit{viz.}, conversion) and the objects involved X and Y and also adds some more relevant details to the summary. In contrast, the ASAP approach misses the objects involved (A, B and C instead of X and Y), although it gets the action of conversion correct.

This theme of identifying the action correctly but not the objects that it operates upon repeats across rows 2 and 3 as well with ASAP. It should be noted that the method name itself had the action in all these cases. Rows 4 and 5 show that in some cases, the ASAP method completely misses the computational intent of the method while our approach does convey the intent. Note that in row 4 our approach is even more precise than the developer's summary as it prefixes \emph{given} to the \emph{T} and \emph{specification} which are both parameters to the method. Typically, method parameters are referred to as ``given'' in method summaries.

Rows 6 and 7 show how our summary conveys more information than both the original developer's summary and the ASAP approach. It uses information obtained from the method body to correctly add additional information in a succinct manner. In row 6, removal is not the only operation but balancing is also done. In row 7 validating is done via checking latitude, longitude and altitude which are mentioned in the summary. The method bodies corresponding to rows 6 and 7 are shown in Listing~\ref{lst:remove} and Listing~\ref{lst:validate}, respectively.

\begin{lstlisting}[language=Java, caption=Method corresponding to row 6 of Table 6. Note that for the `masked' study we replace the method name `remove' with `MASKED'.  ,captionpos=b,frame=lines,label=lst:remove]
public void remove(int node)
{
... 70 lines elided for clarity
    release(node);
    rebalance(parent);
}
\end{lstlisting}

\begin{lstlisting}[language=Java, caption=Method corresponding to row 7 of Table 6. Our approach succinctly captures information from the method body in the summary. ,captionpos=b,frame=lines,label=lst:validate]
public void validate(Geometry g)
{
    g.visit(new GeometryVisitor<Void>)
    {
        public Void visit(Circle c) {
            checkLatitude(c.getY());
            checkLongitude(c.getX());
            checkAltitude(c.getZ());
            return null;
        }
... lines deleted for clarity        
\end{lstlisting}

Observe that a good summary should ideally convey more information than that can be obtained from the method name, parameters and the enclosing class name. The summary generated in rows 6 and 7 of Table~\ref{tab:samples-gen-sum}, achieves precisely this using approximately the same amount of words as used by ASAP i.e., our approach tends to be \emph{concise} yet \emph{precise}.

\begin{table*}[h]
    \centering
    \begin{tabular}{|p{0.15cm}|p{5cm}|p{5cm}|p{5cm}|} \hline 
      \textbf{\#} & \textbf{Ground Truth}  &  \textbf{Our Approach} & \textbf{ASAP} \\ \hline 
      1&
       Converts a {\color{blue}{X}} into a {\color{blue}{Y}}.  & 
       Converts a {\color{blue}{X}} to a {\color{blue}{Y}} , mapping its ID and name to Y's specification. & 
       Converts a {\color{red}{A}}'s {\color{red}{B}} into a {\color{red}{C}}
       \\ \hline
       2&
       Check if sub-dependencies are resolved for an {\color{blue}{XYZ}}& 
       Checks if there are any sub - dependencies under {\color{blue}{XYZ}} that have not been specified in the given map of specifications . & 
       The method returns a boolean value indicating whether the given rule has any dependencies on {\color{red}{ABC}} Specification or {\color{red}{DEF}} Specification \\ \hline
        3&
        Add dependencies for {\color{blue}{X}}& 
        Add dependencies for {\color{blue}{X}} based on rules and entities in XML.&
        Add dependencies for {\color{red}{Y}} \\ \hline
        4 &
        Resolve the {\color{blue}{B}} based on the {\color{blue}{T}} and provided
        specification &
        Resolves an optional {\color{blue}{B}} based on the given {\color{blue}{T}} and specification &
        Get the {\color{red}{X}} from a {\color{red}{Y}}. \\ \hline
        5&
        Reads json files from a zip and creates a map where for each entry the key is the file name and value is the file content
        & 
        Returns a map of JSON file names to their contents from a given ZIP file.&
        Converts a JSON string into an object of type T using Jackson's ObjectMapper \\ \hline
        6 &
        Remove the specified node from the tree. &
        Removes a node from the tree {\color{blue}{while maintaining balance}}.&
        method for removing an element from a binary search tree \\ \hline
        7 &
        Validates the geometry and throws IllegalArgumentException if the geometry is not valid
         &
        Validates the given geometry by checking its {\color{blue}{latitude, longitude, and altitude values using the provided visitor object}}. &
        a validation method that checks if the input geometry object satisfies certain constraints or requirements.  \\ \hline

    \end{tabular}
    \caption{Samples of Generated Summary Comments (X etc are obfuscated identifier names for confidentiality). Method name not masked. Rows 1 to 4, {\color{blue}`blue'} indicates correct objects matching the ground truth, while {\color{red}`red'} indicates incorrect objects. Rows 6, 7, `blue' indicates additional useful information obtained from the body of the method by our approach.}
    \label{tab:samples-gen-sum}
\end{table*}

\subsection{Summary of Summarization Study}
\begin{mdframed}
Our light weight prompt based approach performs as well as or better than the SOTA ASAP, across three different LLMs; across a majority of eight similarity metrics; across, commercial and open source projects; and, across mean, median and obtaining the best scores among the evaluation Java methods.

\end{mdframed}

\begin{mdframed}
Our ablation study of masking the method name during summarization suggests that our approach is less susceptible in a statistically significant way to method name masking than the baseline ASAP approach. This indicates that our approach is more robust to method name choices (idiosyncracies) and perhaps generates the summary from the body of the method.
\end{mdframed}

\subsection{Replication Package} While we cannot divulge Ericsson source code used in our study, we release a replication package consisting of data from the two open source projects \href{https://figshare.com/articles/dataset/replication-package_zip/25164836}{here}.
This has the different prompts, the train and test json files and the results, including the summaries generated by our prompts and the ASAP prompt along with the original developer written summary.

\section{Related Work}
\label{sec:rel-work}
Some of the early works for automatic code summarization at method level are those proposed by Sridhara et al. in ~\cite{giriprasad-summarization-ase10,giriprasad-icse11,giriprasad-icpc-11}. These are based on heuristics which utilize natural language processing and static program analysis. Comment generation at other levels such as exceptions, class, program changes and suggestion of linked APIs (@see also tags), have been described in \cite{summary-class,exceptions-comments,code-changes,at-see-comments}, respectively.

A Deep learning based approach towards code summarization has been desribed in~\cite{iyer2016summarizing}.
Transformer-based, pre-trained models such as CodeBERT and CodeT5 have been used to advance the state of the art in code summarization~\cite{vaswani2017attention,codebert,codet5}.

The advent and popularity of LLMs triggered a boost of interest and new directions in study of many software engineering tasks \cite{sridhara2023chatgpt, liu2024reliability, li2024rewriting, rasnayaka2024empirical, fan2023large}. The work in \cite{sridhara2023chatgpt} studies utility of ChatGPT on ten software engineering tasks. It is one of the early studies post the popularity of ChatGPT and observes promise of LLMs for method name suggestion and log summarization along with other tasks. 

A few-shot training based approach for code summarization has been explored using Codex in the work of \cite{ahmed2022few}; a noteworthy observation includes advantages of code summarization using prompt based few-shot training across projects and within projects as against performance on zero-shot or single-shot training. In \cite{sun2023prompt}, prompt learning framework trains a prompt agent to generate prompts to induce high-quality summary from LLMs (StarCoderBase) for given code snippets. The survey paper, \cite{fan2023large}, systematically categorizes and compiles various Large Language Model based approaches for software engineering tasks under 8 broad categories and includes document summarization and comment generation. The challenges highlighted include the current approaches being largely summary generated using retrieved results; also it highlights the shortcomings of currently used metrics to evaluate the code summarization abilities. 

The various facets involved in comment generation has been captured in the work of \cite{mu2023developer} where intent taxonomy of comments such as what, why, how-to-use, how-is-it-done, properties are considered; a developer-intent driven code comment generation approach utilizes exampler retriever along with encoder-decoder blocks provide the intent-guided selective attention to extract the most relevant information from the semantic representations and generating the intent-aware comments. This in-turn provides the motivation for multi-intent comment generation approach proposed in \cite{geng2024large} utilizing Codex LLMs on Java language datasets. The results indicate that the importance of prompts, re-rankers, and number of examples play a pivotal role in improved performance by utilizing few-shot in-context learning approach. 

The more recent work \cite{asap} builds on \cite{ahmed2022few} towards effective prompts - the ASAP proposes to augment prompts with semantic facts automatically extracted from the source code and introduced into the prompt for few-shot training. The facts extracted and introduced as part of the augmented prompt (along with the code) correspond to repository info, tagged identifiers, the DFG, and the desired (Gold) summary. Our work studies the complementary aspects of prompt; we also introduce only relevant snippet(s) of code to the LLM for code summarization. This is achieved within the prompt in multiple ways: pass over exceptions either through instructions 
to trim from code prompt while also accounting for effect of method name in code summarization. 

\section{Conclusion and Future Work}
\label{sec:conc}
In this work, we described our experiments towards improving readability of (telecom specific) source code at Ericsson. Readability directly impacts program comprehension and thus, the maintainability of the source code.

Our chosen technique of improving readability are to automatically generate comments that succinctly summarize the source code, at the Java method level. Such comments are typically written by developers before the body of the method and are part of the \emph{Javadoc} comments.

We explored an LLM-based SOTA approach, \emph{ASAP} to generate the leading Javadoc summary comments for a Java method. While the ASAP approach has been shown to be reasonably effective on open source projects, we explore its viability in  our commercial telecom domain software. More importantly, ASAP, by design has some overhead in terms of requiring static program analysis of the Java method to summarize, and, requiring the presence of commented methods similar to the input method. 

Therefore, we explored simpler prompts. Our explorations towards relatively simpler prompts show that they work well in practice for our software. We  juxtaposed ASAP, with four other relatively simpler and straightforward approaches relying on variations in the input prompt and the provided context. Our approaches only require the input method to be summarized and \emph{do not} require information from program analysis of the method. Further, we do not require existing exemplars similar to the input method, which is especially useful for newly written methods in our code.

For Javadoc summary comment generation, we found that our simpler \emph{WordRestrict} prompt (approach) performed as well or better than ASAP on 100 non-trivial methods from our Ericsson project.

We repeated the above study on two large and popular open source projects, Guava and  Elasticsearch and found similar encouraging results.
We also performed an ablation study on the effect of the method name in the summary comment generation. We masked the method name and asked the LLMs to generate the summary using the ASAP and our four approaches. Here, we noticed that our approach is
less susceptible in a statistically significant way to method name masking than the baseline (ASAP) approach.

This indicates that our approach is more robust to method name variations and possibly derives
the summary from the body of the method. Thus, our approach may work well even when the developer chosen method name does not precisely capture the computational intent of the method. 


As a part of further investigations, we intend to conduct experiments with other projects and in other languages.
\bibliographystyle{unsrt}  
\bibliography{icsme_2024_arxiv}

\end{document}